\documentclass[aps,twocolumn,amsmath,amssymb]{revtex4}

\usepackage{graphicx}
\usepackage{dcolumn}
\usepackage{bm}

\usepackage{epstopdf}
\begin{document}

\title{Aharonov-Bohm-Like Oscillations in Quantum Hall Corrals}

\author{M.D. Godfrey$^{1}$, P. Jiang$^{1}$, W. Kang$^{1}$, S.H. Simon$^{2}$, K.W. Baldwin$^{2}$, L.N. Pfeiffer$^{2}$, and  K.W. West$^{2}$}

\affiliation{$^{1}$James Franck Institute and Department of Physics,  University of Chicago, Chicago, Illinois 60637\\
 $^{2}$Bell Laboratories, Alcatel-Lucent, 600  Mountain Avenue, Murray Hill, NJ 07974}


\begin{abstract}
Experimental study of quantum Hall corrals reveals
Aharonov-Bohm-Like (ABL) oscillations. Unlike the  Aharonov-Bohm
effect which has a period of one flux quantum, $\Phi_{0}$, the ABL
oscillations possess a flux period of $\Phi_{0}/f$, where $f$ is the
integer number of fully filled Landau levels in the constrictions.
Detection of the ABL oscillations is limited to the low magnetic
field side of the $\nu_{c}$ = 1, 2, 4, 6... integer quantum Hall
plateaus. These oscillations can be understood within the Coulomb
blockade model of quantum Hall interferometers  as forward tunneling
and backscattering,  respectively, through the center island of the
corral  from the bulk and the edge states. The evidence for quantum 
interference is weak and circumstantial.
\end{abstract}

\maketitle

The possibility of realizing topologically protected qubits in the
fractional quantum Hall (FQH) regime has generated considerable
interest in the study of interferometry in quantum Hall
systems\cite{Review}.  The proposed qubits\cite{Review,DasSarma05}
seek to take advantage of the exotic quasiparticle statistics
thought to occur in the FQH states found at fillings $\nu =
5/2$\cite{Willett87,Pan99,Xia04} and $12/5$\cite{Xia04}. Theoretical
studies have shown that these quasiparticles possess a
non-locality, and their exchange statistics are
non-Abelian\cite{Moore91,Read99}. 
The nonlocality can in principle keep qubits free from decoherence, and the use of exchange statistics for computation is thought to be inherently fault tolerant\cite{Review}.
Experimental confirmation of the non-Abelian statistics of the $\nu
= 5/2$ state in principle can be performed with Aharonov-Bohm
interferometry of
quasiparticles\cite{DasSarma05,Stern06,Bonderson06}. In the proposed
experiments, adiabatic transport of a mobile quasiparticle is made
about a localized quasiparticle to probe their mutual exchange
statistics. An even-odd effect with respect to the number of localized
non-Abelian quasiparticles can demonstrate the existence of the
non-Abelian statistics\cite{Stern06,Bonderson06}.

Studies of mesoscopic corrals fabricated from GaAs/AlGaAs
heterostructures have reported Aharonov-Bohm-Like (ABL) interference
in the integer quantum Hall regime\cite{vanwees89,Camino05b}. These
periodic magneto-oscillations have been interpreted either as a
consequence of zero-dimensional states formed as a result of
constructive interference of one-dimensional electron waves
traveling along the edge channels\cite{vanwees89} or as
Aharonov-Bohm interference of edge electrons\cite{Camino05b}.
Similar data in the FQH regime has been interpreted as interference
of Laughlin quasiparticles due to fractional
statistics\cite{Camino05}.

In this paper, study of a quantum Hall interferometer fabricated
from an ultra-high mobility  GaAs/AlGaAs quantum well is reported.
The interferometer features a pair of narrow constrictions through
which  a circular corral is connected to the bulk two-dimensional
electron system. Longitudinal and diagonal magnetoresistances in the
quantum Hall regime  reveal a set of prominent, periodic
oscillations reminiscent of Aharonov-Bohm effect. From the flux
period scaling,  the effective area of the corral and the flux
period of $\Phi_{0}/f$, where $f$ is the integer number of fully
filled Landau levels through the constrictions, have been
self-consistenly determined.  Our findings are mostly in agreement
with the predictions of the Coulomb blockade model of quantum Hall
interferometers proposed by Rosenow and Halperin\cite{Rosenow07}.

\begin{figure}[t]
\includegraphics[width=3.2in]{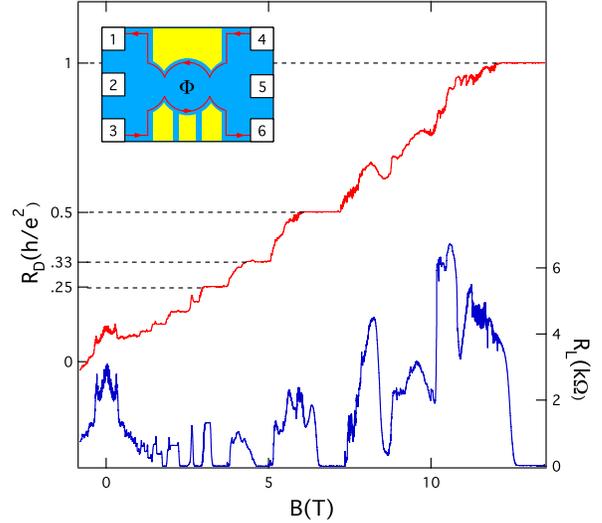}
\caption{Longitudinal, R$_{L}$, and diagonal, R$_{D}$,
magnetoresistance of a quantum corral (T$<10$mK). The inset shows the
layout of the interferometer. The red lines illustrate the edge
state trajectory. Source-drain current is applied between contacts
2-4, the longitudinal magnetoresistance is measured between contacts
3-6, and the diagonal magnetoresistance is  measured between
contacts 1-6.\label{fig:RDRL}}
\end{figure}

The corral was fabricated from a high mobility, symmetrically doped
GaAs/AlGaAs quantum well. The low temperature mobility of the
unprocessed sample was $28.3\times 10^{6}$cm$^{2}$/Vs with an
electron density of $n = 3.3 \times 10^{11}$ cm$^{-2}$. The layout
of the quantum corral is shown in the inset of Fig. 1. The corral
was initally defined using e-beam lithography, dry etched
to the depth of the two-dimensional electron system, and  metallized
with TiAu prior to lift-off.  The diameter of the corral is
1.2$ \mu$m, and the width of the two constrictions is
$\sim$400 nm. The sample was mounted on a dilution refrigerator
capable of reaching below 10 mK. Depletion induced by the etched 
sidewalls reduces  the electron density in the constrictions, $n_{c}$, by
$\sim11\%$ relative to the bulk density, $n_{b}$.
As a result,  $\nu_{c}$, the filling factor across the
constrictions, is smaller than $\nu_{b}$, the bulk filling factor
\cite{Godfrey}. 

\begin{figure}
\includegraphics[width=3.2in]{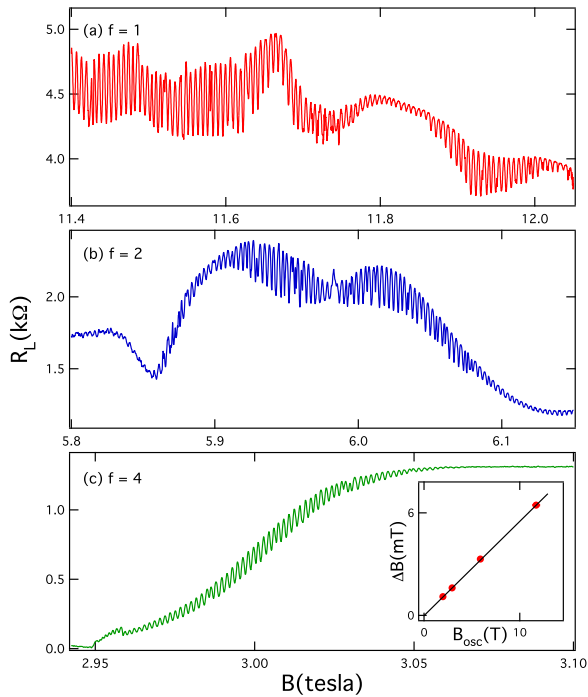}
\caption{Periodic oscillations in the longitudinal
magnetoresistance, R$_{L}$, on the low field side of the (a)
$\nu_{c}$ = 1, (b) $\nu_{c}$ = 2, and (c) $\nu_{c}$ = 4 plateaus.
Inset of (c): Dependence of the  period of ABL oscillations,
$\Delta$B,  on B$_{osc}$, the magnetic field at which they occur.
\label{fig:ABosc}}
\end{figure}

Fig. \ref{fig:RDRL} illustrates the longitudinal, R$_{L}$, and the
diagonal, R$_D$, magnetoresistances of the quantum corral below 10mK
in temperature and under  magnetic field from -1 to 13 tesla.  The
low field ($|$B$| <$ 1 tesla) fluctuations arise as a consequence of
interference of ballistic paths in confined devices whose mean free
path exceeds the dimensions of the device\cite{Marcus}.  Above B $>$
1 tesla, a number of zero resistance regions in R$_{L}$ and plateaus
in R$_{D}$ define the quantum Hall state. As revealed by the R$_{L}$
minima and R$_D$ plateau, strong quantum Hall states are detected at
$\nu_{c}$ = 1, 2, 3, 4, and 6.   From the overlap of the Hall plateaus
in R$_{D}$ and the vanishing R$_{L}$, we find that $\nu_{b} =
\nu_{c}$ overlap occurs for these quantum Hall states.  Such an
overlap is not seen for other quantum hall states as $\nu_{c} \neq
\nu_{b}$ due to density
gradient\cite{LandauerButtiker,Washburn88,Beenakker91}.
Below the $\nu_{c}$ = 1, 2, 4, and 6 plateaus, the ABL oscillations
in R$_L$ and R$_D$ are observed.

Fig. \ref{fig:ABosc} illustrates the ABL oscillations that are
observed in R$_{L}$ on the low field side of the $\nu_{c}$ = 1, 2,
and  4 plateaus, with the ABL oscillations below the $\nu_{c}$ = 1
and 2 plateaus being the strongest.   
The amplitude of the ABL oscillation in R$_{L}$
can be $\ge 15\%$ of the background resistance for $\nu_{c}$ = 1 and
2 ABL oscillations. Depending on temperature and thermal cycling,
over $\ge $300 periods can be detected for a single set of ABL
oscillations.  Interestingly no oscillations are detected on the low
field side of the $\nu_{c}$ = 3, 5, and other odd integer plateaus.
At lower temperatures, ABL oscillations for even $\nu_{c} \le 10$ have been
detected \cite{Godfrey}.

The inset of Fig. \ref{fig:ABosc}c shows the linear dependence of
the ABL oscillation period $\Delta$B as a function of B$_{osc}$, the
center of the magnetic field range at which the periodic
oscillations are found. The values of $\Delta$B and B$_{osc}$ scale
in a way that yields a ratio of
1:$\frac{1}{2}$:$\frac{1}{4}$:$\frac{1}{6}$ between the first four
set of oscillations. Such a scaling of $\Delta$B and B$_{osc}$
always yields the number of fully filled Landau levels, $f$, or
equivalently the quantum number of the quantized Hall plateaus in
R$_{D}$ with the ABL oscillations appearing on the low field side.
Table \ref{tab:table1} summarizes the properties of ABL oscillations
from the data shown in
Figs.\ref{fig:RDRL}-\ref{fig:ABosc}. Similar scaling of $\Delta$B
and B$_{osc}$ has been reproduced under different illumination and
cooldown conditions\cite{Godfrey}.

\begin{table}
\caption{\label{tab:table1} Properties of first 4 largest set of
Aharonov-Bohm-Like oscillations.  $\Delta$B is the
period of oscillations (with variance of $\Delta$B shown for $f$ = 1
and 2 oscillations), B$_{osc}$ is the center of the oscillation
range, $\Delta$B$_{osc}$ is the range of oscillation, $f$ is the integer number of the
fully filled Landau levels in the constrictions, and $r =
\sqrt{\Phi_{0}/\pi f\Delta B}$ is the radius of the area of flux
quantization.}
\begin{ruledtabular}
\begin{tabular}{r|cccc}
$\Delta$B(mT)   & 6.45 $\pm 0.18$ &  3.30 $\pm$ 0.24  & 1.61  & 1.09  \\
 B$_{osc}$(T) & 11.71 & 5.89 & 2.94 & 1.97 \\
$\Delta$B$_{osc}$(T) &   1.20 & 0.60 & 0.16 & 0.06 \\
 $\Delta$B$_{n}$/$\Delta$B$_{1}$ & 1 &  0.51 &  0.25 & 0.17\\
 B$_{osc}^{n}$/B$_{osc}^{1}$  & 1 & 0.50 & 0.25 & 0.17 \\
$1/f$  & 1 & 1/2 & 1/4 & 1/6\\
  $r$(nm) &  452 & 447 & 452 & 449\\
\end{tabular}
\end{ruledtabular}
\end{table}

For a quantum Hall corral of radius $r$, the enclosed flux is $\Phi$
= B$\pi r^{2}$. From the scaling of the magnetic field periods in
Table 1,  the period $\Delta$B of the ABL oscillations is
phenomenologically related to the flux period  by $\Delta B \pi
r^{2} = \Phi_{0}/f$. Shown in Table 1, the radius calculated for
each $\Delta$B consistently yields $\sim$450nm as the radius of the
corral. An active radius of 450nm appears to be reasonable for the
corral with diameter of 1.2$\mu$m once the effect of side-wall
depletion is accounted for.   It is notable that  $\Phi_{0}/f$ is
the flux period even though  the $f + 1^{st}$ Landau
level is partially occupied within the constriction.

\begin{figure}
\includegraphics[width=3.2in]{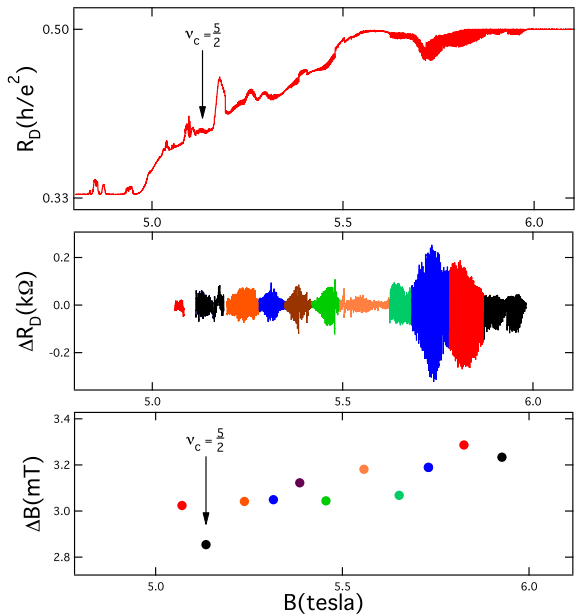}
\caption{(a) Diagonal resistance between $\nu_{c}$ = 2 and 3 under a different 
illumination condition than Fig. 1. (b) Expanded view of the oscillations with 
background resistance subtracted. (c) Oscillation periods in the second Landau level.
\label{fig:plateaus}}
\end{figure}

An intriguing feature of quantum Hall corrals is that it exhibits an
extended sequence of ABL oscillations in the second Landau level.
Fig. \ref{fig:plateaus}a shows R$_{D}$ with ABL oscillations
beginning immediately on the low field side of $\nu_{c}$ = 2 plateau
and lasting until nearly the $\nu_{c}$ = 3  plateau. Fig.
\ref{fig:plateaus}b shows $\Delta$R$_{D}$, which is R$_{D}$ with  a
smooth background subtracted. $\Delta$R$_{D}$ shows apparent
amplitude modulation with the strongest oscillations found
just below the $\nu_{c}$ = 2 plateau. Fig. \ref{fig:plateaus}c shows
the magnetic field period $\Delta$B determined from $\Delta$R$_{D}$
data in  Fig. \ref{fig:plateaus}b. Each period was determined by
Fourier analysis of a different window of magnetic field. 
A reduction of  10\% in $\Delta$B between $\nu_{c}$ = 2 and 3 is found with
some scatter.  Changing periods may mean
that different areas are being probed. Indeed, we see there is a
sudden change where $\nu_{c} \approx 5/2$ where one might expect a
rearrangement of densities in the corral due to the condensation of
the 5/2 state.

\begin{figure}
\includegraphics[width=3.2in]{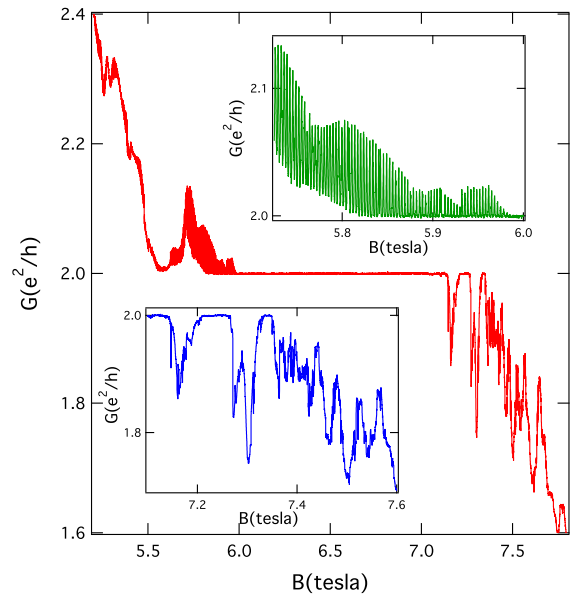}
\caption{Conductance of quantum Hall corral between filling factors
$\nu_{c}=1.6$ and $\nu_{c}=2.4$. Upper (lower) inset: expanded view
of the conductance on the low (high)  magnetic field side of the
$\nu_{c} = 2$ plateau. \label{fig:asymmetry}}
\end{figure}

Fig. \ref{fig:asymmetry} illustrates the striking asymmetry 
 between the low and the  high field sides of $\nu_{c}$ = 2 plateau. 
The conductance $G$ = 1/R$_{D}$ shows a particularly strong set of 
ABL oscillations on the low magnetic field side of the $\nu_{c}$ = 2 plateau. 
The upper inset of Fig. \ref{fig:asymmetry} shows an expanded view of
$G$ on the immediate, low field side of the $\nu_{c}$ = 2 plateau. 
A series of  equally spaced, Coulomb blockade-like
conductance peaks is observed as the magnetic field is reduced from
the $\nu_{c}$ = 2 plateau. 
The lower inset of
Fig. \ref{fig:asymmetry} shows an expanded view of conductance on
the high magnetic field side of the $\nu_{c}$ = 2 plateau.  Unlike
the low field side, there are no ABL oscillations as the conductance
exhibits irregular, jagged dips. Fourier analysis shows 
no dominant oscillation frequency.

Our experiment establishes the following notable features of the ABL
oscillations: (i) ABL oscillations are observed only on the low
magnetic field side of the $\nu_{c}$ = 1, 2, 4, 6, and  other even
integer Hall plateaus. Except below the $\nu_{c}$ = 1 plateau,
ABL oscillations are noticeably absent  below odd integer Hall plateaus.
(ii) No ABL oscillations are observed on the immediate, high field
side of Hall plateaus.  (iii) The ABL oscillations possess a flux
period of $\Phi_{0}/f$ with $f$ is the number of fully filled Landau
levels. The  $f + 1^{st}$ Landau level is partially occupied within
the constrictions. (iv) A typical set of oscillations terminates
within 300 periods or less. (v)  Below the $\nu_{c}$ = 2 plateau,
the ABL oscillations persists over nearly the entire second
Landau level.  (vi) There is a noticeable amplitude modulation and
associated variation of the oscillation period for an extended set
of oscillations. (vii) The ABL oscillations can be detected even
when the constriction is in a compressible state, i.e. there is no
quantized Hall state. This is most clear  below the $\nu_{c}$ = 1
and 2 plateaus.

The features noted above establish that the ABL oscillations are not
the Aharonov-Bohm effect, which possesses a flux period of
$\Phi_{0}$ and two copropagating interference trajectories.
Similarly, the noninteracting model of Ref. \onlinecite{vanwees89}
predicts flux period of $\Phi_0$, and is not consistent with our
results. In contrast, our findings are mostly in agreement with the
Coulomb blockade model of quantum Hall interferometers proposed by
Rosenow and Halperin\cite{Rosenow07}.   In the simplest version of
this model, the constriction consists of an incompressible fluid at
integer filling fraction $f$ and the corral contains a compressible
island at somewhat higher density.  Transport through the system is
effected by (A) forward tunneling from the bulk through the island,
(B) backscattering from the edge state through the island, and (C)
backscattering between edge states via tunneling at the
constrictions.

The first two processes (A and B) occur via tunneling through the
center island and are enhanced when the Coulomb blockade condition
is satisfied, i.e., when the energy of $N$ and $N+1$ electrons on
the island is equal. Since the constrictions are quantized at
filling fraction $f$, addition of a flux quantum $\Phi_0$ transports
$f$ electrons into the central compressible island region,
satisfying the Coulomb blockade condition $f$ times.  Thus these
processes show oscillating conductivity with a magnetic flux period
of $\Phi_{0}/f$.

On the lower magnetic field side of the plateau, the Rosenow-Haperin
model predicts that either forward tunneling from the bulk (A) or
backscattering from the edge (B) should be predominant. The positive
conductance peaks shown in the low field side of $\nu_{c}$ = 2
plateau shown in Fig. \ref{fig:asymmetry} identify each peak as a
forward tunneling event that increases the overall conductance.
Below the $\nu_{c}$ = 1, 2, 4, and 6 plateaus, conductance uniformly
increases, suggesting predominance of forward scattering in ABL
oscillation immediately below these Hall plateaus.  This is quite
natural since on the low field side of the plateau the island and
the bulk are particularly close together\cite{Rosenow07}.

On the high field side of Hall plateaus, the model\cite{Rosenow07}
favors backscattering either via  tunneling through the island (B)
or tunneling at the constrictions (C).   Since backscattering
reduces the overall conductance,  the negative conductance features
found above the $\nu_{c}$ = 2 plateau shown in Fig.
\ref{fig:asymmetry} can be interpreted as a sign of backscattering.
However, the irregularity of the signal prevents us from making a
clean identification of which process is involved.  In particular,
we cannot identify any particular region as being due to
interferences associated with process (C).  It is, of course,
possible that the complicated and irregular nature of the signal is
a result of several competing processes.

A salient feature of the ABL oscillations is that it is detected
below the $\nu_{c}$ = 1, 2, 4, 6 and other even integer Hall
plateaus, but not below other odd plateaus.   Clearly, spin plays a
nontrivial role in whether or not ABL oscillations can be observed.
This is likely because the smaller exchange gap for odd integers (compared with the cyclotron gap of even integers) makes the physical distance between the 2n-1 and 2n edges smaller than that between the 2n and 2n+1 edges. Thus, in the
forward scattering process (A) the distance required for a particle
to tunnel from the bulk to the island could be smaller on the low
field side of an even plateau than on the low field side of an odd
plateau.  We note, however, that for process (B) such an even-odd
effect would not seem likely as long as the tunneling from the edge
to the bulk preserves spin.

The apparent modulation of the intensity of the ABL oscillations
could be from any one of a number of sources.   Such modulation
could caused by slow changes in the overlap of wavefunctions
involved in the relevant tunneling process, by slow rearrangement of
densities, or by beating between competing tunneling processes.

In contrast to the
model described by
Rosenow-Halperin\cite{Rosenow07}, we detect ABL oscillations when
the filling factors inside the constrictions are in a compressible
regime.   Indeed, for the ABL below $\nu=2$ roughly the same period
oscillations are seen almost all the way to $\nu=3$.   While this is
slightly outside of the orthodox model, it allows us to conclude
that throughout this region: (a) The tunneling is always single
electron tunneling. (b) The oscillations are from a Coulomb blockade
origin. (c) Addition of a flux quantum $\Phi_0$ continues to add
an integer number of $f$ electrons to the island, even though the
constriction may have a partially filled $f+1^{st}$ landau level.
This result is quite interesting, since in a sample of this high
mobility, one might expect FQH states in the
constrictions.  Indeed, (as shown in Fig.~\ref{fig:plateaus}) there
appears to be an effect in the flux period when the constrictions have
the density of the FQH state (although evidence for
this is somewhat weak). If the constriction is really quantized at
filling $\nu_c$, one might expect that addition of a flux quantum
adds charge $\nu_{c}e$. However, this does not appear to be the case.

In summary, we have studied the ABL oscillations in quantum Hall
corrals. The ABL oscillations are most prominent in the low field
side of the $\nu$ = 1, 2, 4, 6... quantum Hall plateaus. They can be
detected over extended ranges of magnetic fields, including over the
compressible filling factors. Their flux periods is $\Phi_{0}/f$,
where $f$ is the integer number of fully filled Landau levels
through the constriction. These features establish that ABL
oscillations do not arise from Aharonov-Bohm effect. Instead, these
oscillations can be identified as electron tunneling peaks due to
Coulomb blockade through the island in the corral. It will be
necessary to distinguish Coulomb blockade effects from interference
effects in the future.

We thank N. Cooper, F.D.M. Haldane, N.P. Ong, S. Sondhi, D.C. Tsui, and A. Yazdani  for useful discussions. This work is supported by the Microsoft Q Project and the University of Chicago MRSEC. W.K. acknowledges  Guggenheim Foundation for support and Princeton University for hospitality during his stay.


\begin{thebibliography}{99}
\bibitem{LandauerButtiker}  Analysis of the transport through the quantum corral within the Landauer-Buttiker formalism\cite{Washburn88,Beenakker91} yields a longitudinal magnetoresistance $R_{L}  =  \left( \frac{1}{\nu_{c}} - \frac{1}{\nu_{b}}  \right)  \frac{h}{e^{2}}$ and diagonal resistance  $R_{D}  =  \frac{1}{\nu_{c}}  \frac{h}{e^{2}}$, where $\nu_{c}$ and $\nu_{b}$ are the filling factors for the two-dimensional electrons at the constrictions and bulk, respectively. Results of detailed study
will be reported elsewhere\cite{Godfrey}
\end{thebibliography}
\end{document}